**Two accreting proto-planets around the young star PDS 70**

S. Y. Haffert[1*], A. J. Bohn[1], J. de Boer[1], I. A. G. Snellen[1], J. Brinchmann[1,2], J. H. Girard[3,4], C. U. Keller[1], R. Bacon[5]

1 Leiden Observatory, Leiden University, Postbus 9513, 2300 RA Leiden, The Netherlands
2 Instituto de Astrofísica e Ciências do Espaço, Universidade do Porto, CAUP, Rua das Estrelas, PT4150-762 Porto, Portugal
3 Space Telescope Science Institute, 3700 San Martin Drive, Baltimore, MD 21218, USA
4 Université Grenoble Alpes, CNRS, IPAG, 38000 Grenoble, France
5 CRAL, Observatoire de Lyon, CNRS, Université Lyon 1, 9 Avenue Ch. André, F-69561 Saint Genis Laval Cedex, France

**Newly forming proto-planets are expected to create cavities and substructures in young, gas-rich proto-planetary disks [1-3], but they are difficult to detect as they could be confused with disk features affected by advanced image-analysis techniques[4,5]. Recently, a planet was discovered inside the gap of the transitional disk of the T-Tauri star PDS 70[6,7]. Here we report on the detection of strong H-alpha emission from two distinct locations in the PDS 70 system, one corresponding to the previously discovered planet PDS 70 b, which confirms the earlier Hα detection[8], and another located close to the outer-edge of the gap, coinciding with a previously identified bright dust spot in the disk and with a small opening in a ring of molecular emission[6,7,9]. We identify this second Hα peak as a second proto-planet in the PDS 70 system. The Hα emission spectra of both proto-planets indicate ongoing accretion onto the proto-planets[10,11], which appear to be near a 2:1 mean motion resonance. Our observations show that adaptive-optics-assisted, medium-resolution, integral-field spectroscopy with MUSE[12] targeting accretion signatures will be a powerful way to trace ongoing planet formation in transitional disks at different stages of their evolution. Finding more young planetary systems in mean motion resonance would give credibility to the Grand Tack hypothesis in which Jupiter and Saturn migrated in a resonance orbit during the early formation period of our Solar System[13].**

PDS 70 (V* V1032 Cen) is a young T-tauri star at a distance of 113.43+-0.52 pc [14,15] with a spectroscopically determined age of 5.4+-1.0 Myr [7]. Its proto-planetary disk was first discovered through spectral energy distribution(SED) modelling [16], and later directly imaged at near-infrared and sub-mm wavelengths [9,17,18]. Both the SED modelling and direct imaging show that PDS 70 harbours a transitional disk in which a large radial region from 20 AU – 40 AU[6, 18], as seen in the near-infrared, is cleared of dust. The gap is found to be larger at sub-mm wavelengths [6,18], indicating that there is a radial segregation of dust grains possibly generated by a radial pressure gradient in the disk. A potential mechanism for such a pressure bump is the dynamical clearing of the disk by the planetary-mass companion PDS 70 b[19,20]. PDS 70 b is estimated to be a 4-17 $M_{Jup}$ planet at a projected angular separation of 195 milli-arcseconds (22 AU) and position angle of about 155°. Follow-up imaging observations with MagAO on the Magellan 6.5m telescope revealed a tentative 4σ Hydrogen α(Hα) emission signal [8]. Hα emission is commonly associated with accretion[10,11] and therefore indicates that PDS 70 b is still in formation.

The PDS 70 system was observed with the Multi Unit Spectroscopic Explorer (MUSE)[12] at ESO's Very Large Telescope (VLT) in the Laser Tomography Adaptive Optics (LTAO) assisted narrow-field mode during the commissioning night of 20 June 2018. MUSE is an optical, medium-resolution, integral-field spectrograph that covers 480-930nm at a spectral resolving power ranging from $\lambda/\Delta\lambda$ = 1740 in the blue to 3450 in the red, targeting several important accretion tracing emission lines including Hα at λ=656.28 nm. We processed the data by applying a high-resolution Spectral Differential Imaging (HRSDI) technique, in which a scaled, continuum-normalized stellar spectrum is subtracted from the data at each spatial pixel (spaxel) position, and further residuals in the spectrum of each spaxel are removed using a principle component analysis (see Methods for a detailed description). HRSDI is only sensitive to sharp spectral features, such as spectral lines, because the continuum-normalization step removes all differential broadband features between different spaxels. Excess continuum light from starlight that is reflected by the disk around PDS 70 and the continuum of potential planets is also removed. With HRSDI we can side-step the common issues that conventional high-contrast imaging techniques like Angular Differential Imaging(ADI) have, where the spatial information is used to build up a stellar reference. Such techniques are sensitive to the spatial structure of the source and can generate artifacts that look like a planet[4,5].

Using HRSDI we detect a strong Hα emission line at two distinct locations in the PDS 70 system (Figure 1, left panel), with a 11σ source at the position of PDS 70 b, and a second, 8σ source detected at 240 mas away from the host star at a position angle of 283°. The former detection is an independent confirmation of the earlier observations of PDS 70 b[6-8]. The latter source coincides with a previously identified bright spot in 2.1 and 3.8-μm broad-band images [6], which is confirmed by our re-analysis of the 2.1 and 3.8 μm data from SPHERE and NACO at the VLT (Figure 2, middle and right panel).

Gaussian profiles fitted to the extracted emission spectra (see Figure 1) show that the Hα lines from both PDS 70 b and the other source are red-shifted with respect to the stellar Hα emission. This line-of-sight shift is measured with respect to the stellar Hα line before the removal of the starlight; every spaxel is therefore self-calibrated. The Hα line-of-sight redshift for PDS 70 b is 25 +- 8 km s$^{-1}$, and 30 +- 9 km s$^{-1}$ for the second source. The full width at half maximum (FWHM) of the two emission lines are 123 +- 13 km s$^{-1}$ and 102 +- 19 km s$^{-1}$ for PDS 70 b and the other source, respectively. These lines are significantly narrower than the stellar line, which has a width of 147 +- 5 km s$^{-1}$. A third difference between the emission lines are their shapes. The Hα line of the star exhibits an inverse P Cygni profile with a blue-shifted emission and a red-shifted absorption component, which is quite common and a sign of magnetospheric accretion[21,22]. The Hα line profile from PDS 70 b and the other source do not exhibit these features, and show just a single Gaussian-like profile.

The fact that the line shape, velocity offset, and the line widths differ from what is expected from reflected star-light excludes reflected light as the origin and indicates that the line emission is locally generated. The multi-epoch astrometry (see Table 1) of the new point source reveals that it orbits in the direction of the Keplerian rotation of the circumstellar disk and is co-moving with its host star, which makes it improbable to be a background object. We therefore identify this second source of Hα emission as the location of a second, forming planet, PDS 70 c. The significant difference between the line-of-sight radial velocity of Hα emission from the accreting proto-planets compared to their Keplerian velocities, 4.3 km s$^{-1}$ and 3.4 km s$^{-1}$ for PDS 70 b and c, indicates that the dynamics of planetary accretion are different from the orbital dynamics of the planets. This is also the case for accreting stars, where the line-of-sight velocity of the accreting material is different from the system velocity.

The K-L color and non-detection in H-band imply that PDS 70 c is redder than PDS 70 b, which could be either due to a lower temperature and mass or due to obscuration by dust from the disk. Comparing the K-L color of PDS 70 c to evolutionary models indicates that the planet has a mass within the range of 4-12 $M_J$ after having included possible biases from the circumstellar disk and extinction (see Methods section). There is a high likelihood that the estimated mass of the companion, even by including some of the biases, is overestimated due to possible other structures around the planet (see Figure 3) like additional scattered star light, a circumplanetary disk or potential streamers that connect the planet to the circumstellar disk. Detailed hydrodynamical modelling of the system may put a better constraint on the planet masses.

We estimated the mass accretion rates of both companions using an empirical relation between the Hα 10-percent width, which is the full width at 10-percent of the maximum, and the mass accretion rate [10, 11]. The derived 10-percent widths are 224+-24 km s$^{-1}$ and 186+-35 km s$^{-1}$ for planets b and c, which indicate a mass accretion rate of $2 \times 10^{-8+-0.4}$ $M_J$ yr$^{-1}$ for b and $1 \times 10^{-8+-0.4}$ $M_J$ yr$^{-1}$ for c. The combined mass accretion rates of the planets are comparable to the stellar accretion rate, which is $5.5 \times 10^{-8+-0.4}$ $M_J$ yr$^{-1}$. A previous estimate of the accretion rate of planet b, based on Hα the luminosity [8], estimated the mass accretion of PDS 70 b to be $1 \times 10^{-8+-1}$ $M_J$ yr$^{-1}$. The estimate of [8] based on the absolute Hα luminosity needs to take the extinction into account, which was chosen as 3.0 magnitudes to include the effects of a possible circum-planetary disk and the extinction by the circum-stellar disk. Their measurement is consistent with the low accretion rate found here, which is not affected by extinction as it is based on the width of the Hα line instead of the luminosity. At the current accretion rate, it would take 50-100 Myr to form Jupiter-mass planets, which is much longer than the typical disk lifetime of 10 Myr. The accretion rates of young stars are significantly decreasing near the end of their formation when they exhibit episodic accretion bursts [23]. A similar scenario of long-term variable accretion might apply to the planets in this system. Short term variability is due the accreting gas dynamics, which for example could be a rotating hotspot on the planet surface or an orbiting accretion funnel from the circum-planetary disk.

Based on the wide gap in the disk and the low accretion rate of the star, PDS 70 was already predicted to contain a multi-planet system[18] where the planets block the mass flow towards the star. Hydrodynamical modelling of viscous disks suggests that a single, massive planet is only able to open a small gap (<15 AU) [1,2]. In contrast, multi-planet systems are thought to be able to carve large gaps in disks with azimuthal asymmetries [2,3], where the gap width is governed by the dynamical and viscous time scales of the disk. Detailed hydrodynamical modelling will be necessary to constrain the disk-planet interaction of the PDS 70 system. Another indication of a planet at a larger orbital distance came from the 870-µm ALMA observation of the HCO+ molecule that shows a hole in the disk structure, while the dust continuum does not show this[9]. We suspect that this is caused by PDS 70 c since the location of the hole and PDS 70 c coincide.

Assuming that the orbits of the planets and the disk are coplanar (at an inclination of 49.7º [18]), angular separations to the host star imply orbital distances for PDS 70 b and c of 20.6+-1.2 AU (consistent within the 1σ uncertainties of earlier measurements[6,7]) and 34.5+-2 AU, respectively, suggesting that the planets are in or near a 2:1 mean motion resonance if the orbits are circular. Mean motion resonance migration is proposed as an early orbital evolution scenario for massive gas giant planets[24]. During a migration-II scenario, in which planets are massive enough to carve a gap in the disk, the inward migration speed depends inversely on the mass of the planet[25]. If the inner planet is more massive than the outer, the latter will migrate towards the inner planet until it is captured in resonance, locking their relative orbits[26]. A specific version of such an orbital evolution scenario is suggested for the gas giant planets in the early solar system as the Grand Tack Hypothesis[13]. While the most massive planet Jupiter was formed first and was migrating towards the young Sun, the formation and subsequent faster inward migration of Saturn locked the two planets in a 3:2 mean motion resonance. This is thought to have reversed their, now coupled, migration direction outward towards their current orbits. Finding other systems, like PDS 70, in mean-motion resonance lends credibility to this formation scenario for our solar system.

The observations with MUSE show that adaptive-optics assisted, medium-resolution, integral-field spectrographs are highly efficient instruments to observe accretion emission spectra from planets in formation. We suspect that more accreting planets could be found and characterized in transition disks with this technique. Another exciting opportunity, which is now possible due to the high efficiency of both the HRSDI data processing technique and MUSE, is the detailed investigation of accretion on short and long time scales. This will shed light on the variability of planetary accretion and will lead to a better understanding of the formation of planets, allowing us to infer the formation history of the solar system.

**Acknowledgements**
Based on observations collected at the European Organisation for Astronomical Research in the Southern Hemisphere under ESO programmes 60.A-9100(K), 095.C-0298(A), 097.C-0206(A) and 097.C-1001(A). A.B. and J. de Boer acknowledge support the ERC Starting Grant 678194 (FALCONER). I.S. acknowledges funding from the European Research Council (ERC) under the European Union's Horizon 2020 research and innovation program under grant agreement No 694513. R.B. acknowledge support from the ERC advanced grant 339659-MUSICOS. JB acknowledges support by FCT/MCTES through national funds by this grant UID/FIS/04434/2019 and through the Investigador FCT Contract No. IF/01654/2014/CP1215/CT0003. The authors acknowledge the ESO AOF and Paranal teams for their expertise and support during the commissioning activities.

**Author Contributions**
This work made use of the newly commissioned narrow-field mode of MUSE on the VLT which was led by R. Bacon. S. Haffert, J de Boer, J. Girard and J. Brinchmann prepared the observations, which were executed by R. Bacon and J. Brinchmann.  J. Brinchmann reduced the raw data products from MUSE with the ESO MUSE pipeline.  S. Haffert wrote the HRSDI pipeline, processed, analyzed the combined data cubes, performed the astrometry and orbit fitting. A. Bohn reduced the archival SPHERE and NACO data, performed the photometry and astrometry on these data sets and wrote the corresponding sections. I. Snellen and C. Keller supervised the effort of S. Haffert. All authors contributed to key aspects of the manuscript.

**Author Information**

The authors declare no competing financial interests. Correspondence and requests for materials should be addressed to S. Y. Haffert (haffert@strw.leidenuniv.nl).

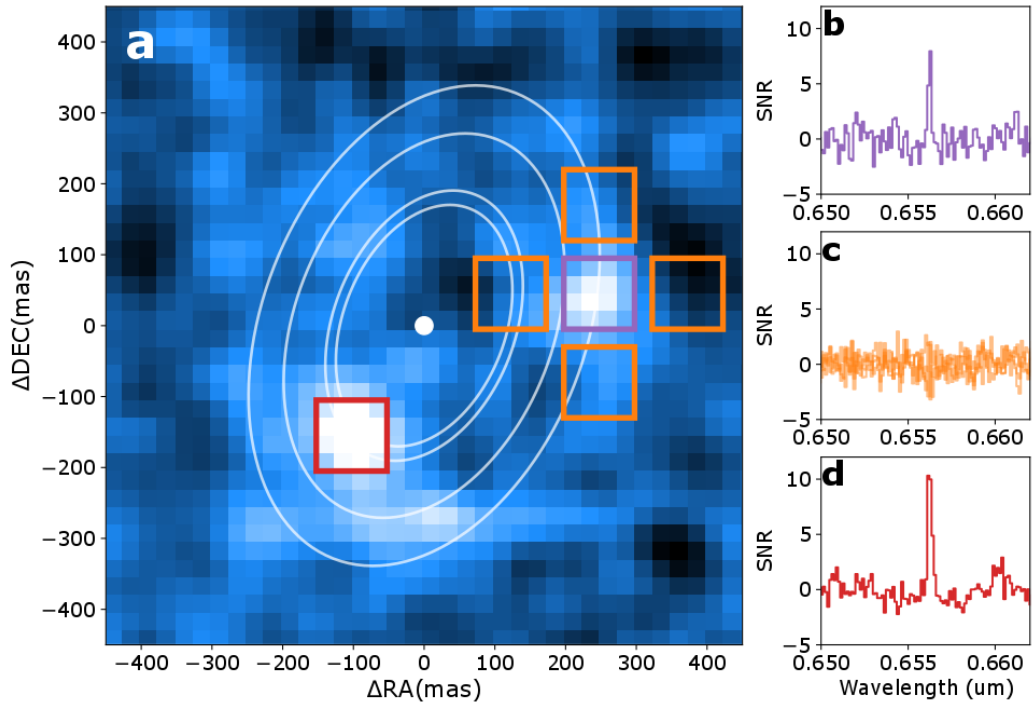

**Fig. 1 Overview of the PDS 70 system. a,** The Hα detection map with an overlay of the contours of the orbital radii and a white dot in the center that marks the position of the star. The contours for PDS 70 c are the minimum and maximum orbital radii found for the different wavelength observations. For both objects the square apertures that were used for the photometry are shown, with the red aperture for PDS 70 b and the purple aperture for PDS 70 c. **b, c, d,** The corresponding spectra divided by their standard deviation are on the right and centered around the Hα line position. The four apertures in orange indicate reference areas that are used to compare with PDS 70 c. The orange reference spectra on the right do not show any spectral feature, while both PDS 70 b and c clearly show Hα in emission.

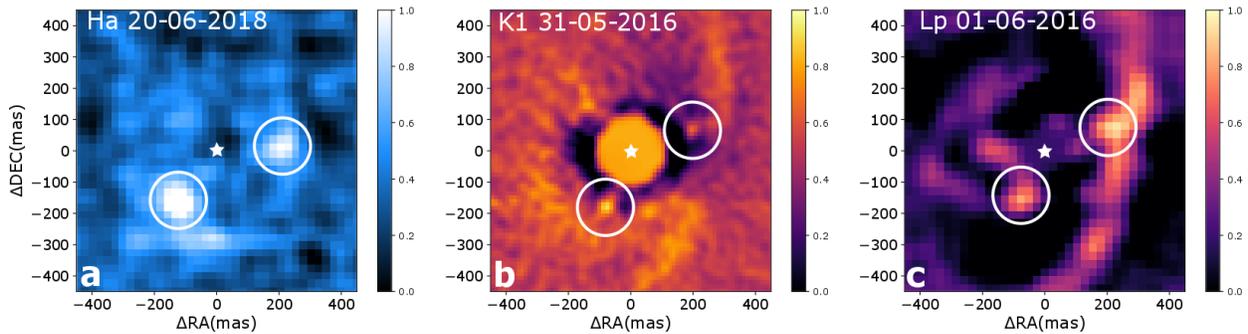

**Fig. 2 Multi-epoch detection of PDS 70 b and c. a**, The Hα detection map of PDS 70 after removal of all direct and scattered starlight. PDS 70 b and c have their position marked with white circles. The white star in the middle shows the position of the star. The feature just south of PDS 70 b is most likely caused by image slicing in MUSE, as the signal is perfectly aligned with the slicing and field splitting axis and we do not see it in the other datasets. **b**, The K1-band observations of PDS 70 with SPHERE/IRDIS after Angular Differential Imaging processing. Both companions, with their positions marked by the white circles, are recovered in the data. The position of the star is marked by the white star in the center. **C**, NACO observations in L-band. PDS 70 b is clearly visible, while PDS 70 c is connected to the disk. This is due to the large full width half maximum of the PSF in L-band. The disk should be a smooth ring according to previous research, the extension at the position of PDS 70 c is therefore most likely due to the planet itself. The position of the star is marked by the white star in the center.

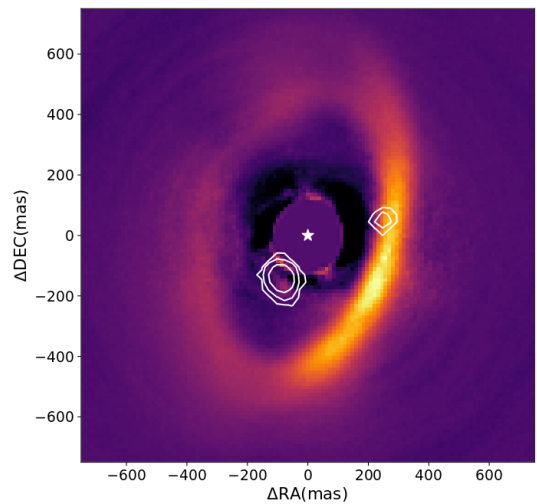

**Fig 3. Composite image of the proto-planetary disk and the two companions.** The disk image from SPHERE H-band observations is overlaid with the contours of the Hα emission of the two planets to indicate their positions in the disk. The disk features a smooth outer disk and a perturbed inner disk. The darker features around the inner disk are due to over-subtraction of the inner-disk by the post-processing algorithm. The planets are clearing out the large gap between the outer and inner disk edges, which can be seen in the scattered light image. The outer planet is partially obscured by the outer disk.

**Table 1 The photometry and astrometry of PDS 70 b and c.**

| Component | Date | Instrument | Band | ΔM (mag) | Line flux (1E-16 erg/s/cm^2) | PA(°) | Separation (mas) | Orbital radius (AU) |
|---|---|---|---|---|---|---|---|---|
| PDS 70 b | 2016-05-14 | SPHERE | K1 | 8.0 +- 0.2 | | 152.4 +- 1.5 | 186.13 +- 7 | 21.3 +-1 |
| PDS 70 b | 2016-06-01 | NACO | L' | 6.9+-0.25 | | 151.4 +- 2.0 | 181.2 +- 10 | 22.2 +- 3 |
| PDS 70 b | 2018-06-20 | MUSE | Hα | **6.9+-0.1** | **3.9 +- 0.37** | 146.8 +- 8.5 | 176.8 +- 25 | 21.0 +- 3 |
| PDS 70 c | 2016-05-14 | SPHERE | K1 | 8.8+-0.2 | | 285 +- 1.5 | 215.1 +- 7 | 33.6 +- 1.5 |
| PDS 70 c | 2016-06-01 | NACO | L' | 6.6+-0.2 | | 283.3 +- 2 | 254.1 +- 10 | 39.7 +- 4.5 |
| PDS 70 c | 2018-06-20 | MUSE | Hα | **7.7+-0.2** | **1.9 +- 0.32** | 277.0 +- 6.5 | 235.5 +- 25 | 38.6 +- 4.5 |

**Methods**

**VLT/MUSE observations and data reduction.** The PDS 70 system was observed during the commissioning run of the adaptive optics assisted MUSE narrow-field mode in June 2018. MUSE is a medium resolution integral field spectrograph (IFS), which acquires a spectrum for every spatial pixel in a continuous field of view of 7.5"x7.5"[12]. It is coupled with GALACSI, the advanced laser tomography adaptive optics system of the VLT Adaptive Optics Facility [27,28], which was also used during the observations of PDS 70. PDS 70 itself was used as the low-order tip-tilt star. The observations were split into six individual exposures of 300 seconds each in field stabilizing mode with a total of 1800 seconds on target. Between each observation the field of view was rotated by 90 degrees. This was done to reduce the impact of possible bad pixels, reduce flatfield errors and improve the sampling of the line spread function. Each observation was processed with version of 2.4 of the MUSE pipeline which is available through ESO's web pages[29]. The pipeline performs background subtraction, flat fielding for non-uniform pixel response, spectral extraction and the wavelength calibration. Spatial offsets between the individual exposures were determined manually by finding the centroid position of the star. After the extraction the individual data cubes were stacked to create the combined data cube from all observations.

**High-resolution spectral differential imaging (HRSDI).** We subtracted a reference stellar spectrum from every spatial pixel to find the faint companions. A stellar model could in principle be used to subtract out the star, but this will leave strong residuals either due to intrinsic variability of the host star or due to uncalibrated instrumental effects. We therefore used a data-driven approach to create the reference spectrum, similar to the approach that was used for the characterization of Beta Pictoris b[30]. The changes in the spectrum between spatial pixels can be separated into two parts, low-order effects and high-order effects. The low-order effects change the continuum profile, while the high-order effects change the spectral lines.

To calibrate the low-order effects, we first normalized all spectra by their total flux and created a reference spectrum by taking the median over all spatial pixels. The median is a robust estimator for the mean in the presence of outliers. As the exoplanets are only present in a few spatial pixels, the median will lower the influence of the planet spectra on the reference spectrum.

The spectrum for each individual spatial pixel is then divided by this reference spectrum. Assuming that no high-order effects are present, we will be left with a differential low order continuum signal. This residual is low-passed filtered by applying a first-order Savitzky-Golay filter with a window width of 101 pixels (126.25 Å). Each spectrum is divided by its low-passed differential continuum to correct for the low order residuals.

The high-order residuals are due to several instrumental effects, such as a changing wavelength calibration or a changing resolving power across the field. Because MUSE is an IFS with 310 by 310 spatial pixels, we acquired many realizations of the high-order instrumental effects. We used these to build a reference library through principle component analysis (PCA). Each PCA component will remove different correlated residuals.  PCA was applied over the spectral direction where the first few components corrected for differential wavelength solutions and the subsequent components mainly correct regions that are strongly contaminated by telluric absorption (Oxygen A and B bands). We found the optimum number of components to subtract to be 5 by maximizing the signal-to-noise of PDS 70 b. Only after subtracting on the order of 20 components did the signal from PDS 70 b start to decrease.

The HRSDI method is only sensitive to sharp spectral features, such as spectral lines, because the continuum-normalization step removes all differential broadband features between different spaxels. Excess continuum light from starlight that is reflected by the disk around PDS 70 and the continuum of potential planets is also removed. With HRSDI we can side-step the common issues that conventional high-contrast imaging techniques like Angular Differential Imaging(ADI) have where the spatial information is used to build up a stellar reference. Such techniques are sensitive to spatial structure of the source and can generate features that look like a planet[9,10].

**Aperture photometry of both companions and SNR determination.** We estimated the relative flux by applying aperture photometry to every wavelength channel. Square apertures of 100 by 100 milli-arcseconds (4 by 4 pixels) were chosen because the MUSE PSF during the observations had a full width half maximum of 60-80 milliarcseconds at 656 nm. No background correction for the aperture photometry was necessary as this was already done during the pre-processing. A major concern for high-contrast image processing techniques is the subtraction of the planet signal during the post-processing, which will lead to strongly biased photometry of the companions[31]. We injected fake sources into the dataset to determine how much the HRSDI method would influence the photometry. The recovered signals of the fake sources were within a few percent of the injected signal, indicating that the HRSDI did not influence the photometry. We estimated the error in the photometry by taking the standard deviation along the spectral direction for each region.

The extracted one-dimensional spectra were sub-sequentially used to fit spectra line profiles with the Markov chain Monte Carlo(MCMC) sampler emcee[32]. The Hα emission line was modelled by a Gaussian line profile on top of a constant continuum. To determine the radial velocity reference we fitted a PHOENIX model with an effective temperature of 4000 K and a log(g) of 4.5 [12] at the same spatial positions but before removing the star. From these two measurements we could determine the relative radial velocity of the planets. The uncertainties of the measurements were derived from the posterior distributions from the MCMC procedure (see Supplementary Figures 1 and 2). The ratio between the line flux of the planets compared to the stellar continuum around the Hα line was used to define the contrast ratio.

**Astrometry of the Hα emission from PDS 70 b and c.** We fitted a Moffat profile to the cleaned Hα data cubes. The initial starting point of the fitting procedure was estimated from the Hα image. We used a least-squares approach to fit the astrometry, and we estimated the errors by analyzing the covariance matrix around the optimal $\chi^2$ value. The retrieved error bars from the least-square method were less than one pixels. This is not realistic because the PSF of MUSE is not Nyquist-sampled by the image slicer, and the image slicer causes strong diffraction effects when part of the light hits an edge of any of the slicer mirrors. The effects becomes more apparent when the AO is delivering higher quality images. This makes it difficult to achieve sub-pixel astrometry without a good PSF model, and we therefore used the size of a single spatial pixel as the astrometric uncertainty. We verified the astrometric solution with a background star that was in the field-of-view. This star is roughly 2.3 arcseconds to the north. We can use the astrometric measurements of GAIA[14,15], because PDS 70 and the background star were observed with GAIA. After correction for the proper motion of the two stars, we measure a 0.3 percent difference in the angular separation of the two stars between GAIA and MUSE and a 0.76 degree offset from true north. Using the astrometry points from the SPHERE/NACO and MUSE observations we exclude both planets as background objects (See Supplementary Figure 3).

**Orbit radius and mean motion resonance estimation.** While the L and K-band data were taken on different nights, they are too close in time (1 night apart) to count as different epochs. A minimum of 3 epochs are necessary to fit all parameters of a Keplerian orbit, and we only have two (Hα and K+L-band). Under the assumption of co-planarity with the protoplanetary disk and a circular Keplerian orbit, we determined the orbital radius for both companions. We used earlier measurements of the disk inclination and position angle, which are 49.7° and 158.6° respectively[17,18], to deproject the apparent positions of the planets onto the circumstellar disk plane. After the deprojection we used a weighted mean to calculate the orbital radius for both planets. With the orbital radii and under the assumption of Keplerian motion we measured the ratio of the orbital periods to be 2.0+-0.15, 2.5+-0.25 and 2.6+-0.75 for the K-band, L-band and Hα. In L-band the astrometry is biased to longer orbital radii, due to the close proximity of PDS 70 c to the disk. The measurements suggest a mean motion resonance close to 2:1. To account for the astrometric errors we used a parametric bootstrap method, with a Gaussian error model to estimate the probability distribution of the period ratio (see Supplementary Figure 4).

**SPHERE and NACO archival data reduction.** We obtained three archival datasets on PDS 70 including raw calibration files from the ESO archive: (i) Coronagraphic SPHERE/IRDIS data (PI: Beuzit, ESO ID: 095.C-0298(A)) taken with the H23 dual-band filter combination [33], (ii) non-coronagraphic SPHERE/IRDIS (PI: Matter, ESO ID: 097.C-1001(A)) data taken with the K12 dual-band filter combination, and (iii) non-coronagraphic NACO L' data (PI: Launhardt, ESO ID: 097.C-0206(A)).

The data reduction was performed with our own image processing pipeline, which is based on the new release of the PynPoint package [34]. This includes basic steps such as dark subtraction, flat-fielding, and correction of bad detector pixels with a 5 sigma box filtering algorithm. The NACO data were obtained with a dither pattern, placing the star at three offset positions on the detector. Because one quadrant of the detector showed strong, non-static readout artefacts, we discarded every third cube of the NACO data. For the SPHERE data, we performed the subtraction of the sky and instrumental background using the additional images taken at an offset sky position without any source. Due to the large offsets between the NACO dither positions, we subtracted the thermal background from the science images themselves. We defined regions of interest around the star for each of the two dither positions and obtained a stack of science frames and background frames for each of the two positions. Then, we modelled and subtracted the background in each science frame with an approach based on PCA of the corresponding background stack as described in [35]. Afterwards, the resulting background-subtracted science frames were merged again.

Both SPHERE data in K1-band and NACO data were obtained without a coronagraph and without saturating the detector. Therefore, we were able to center the frames on PDS 70 with a two-dimensional Gaussian fit. The SPHERE data in the H2- band was obtained with an apodized-pupil Lyot coronagraph [36] in place. To center the science images, we made use of the center frames that were taken alongside the science observations. For these frames, a sinusoidal shape was applied to the deformable mirror to create four calibration spots around the star's position behind the coronagraph. These spots were used for the centering procedure, following the instructions in the SPHERE manual.

To detect and to reject bad science frames, we measured the flux inside circular apertures with a radius of 3 FWHMs of the instrumental PSF around the star for the non-coronagraphic data. For the coronagraphic data, we used circular annuli around the coronagraphic mask with inner and outer radii of 100 mas and 160 mas instead. Finally, we discarded all frames deviating from the average aperture flux by more than 3 sigma. After this pre-processing, 60, 672, and 12346 centred images remained for SPHERE H23-, SPHERE K12-, and NACO L', respectively.

Since the data were taken in pupil-stabilized mode, a PSF subtraction based on angular differential imaging [37] in combination with PCA [38] was applied. To model the stellar PSF, we fitted 1 and 20 principal components for SPHERE K12 and NACO data, respectively. The number of fitted components were optimised in order to maximise the signal-to-noise ratio of PDS 70 b.

Because ADI-based algorithms tend to produce unphysical residuals when applying them to extended structures such as disks [39], we used an approach based on reference star differential imaging (RDI, [40]) in combination with PCA to reduce the coronagraphic SPHERE data instead. We chose the K9IVe-star TYC 7433-1102-1 as a reference, which was observed during the same night (2015-05-03) with the exact same observational setup (PI: Beuzit, ESO ID: 095.C-0298) as the data on PDS 70. The pre-processing was performed analogously to the reduction of the coronagraphic SPHERE data described above. After frame rejection, we were left with 63 centred reference images. These were analysed with respect to their principal components and 5 components were fitted to the pre-processed science images of PDS 70 to model the stellar halo and the instrumental artefacts. The PSF models were subtracted from the science images, the residual frames were rotated such that north is up and east is left, and the remaining stack was median combined. We optimised the number of fitted principal components to maximise the signal-to noise ratio of the disk around PDS 70 in the final image.

**Astrometry and photometry extraction of PDS 70 b and c from NACO and SPHERE data**. The astrometry and photometry of each companion were extracted simultaneously by injecting artificial fake companions using the *SimplexMinimizationModule* of the new *PynPoint* release [38]. For the astrometric extraction we use plate scales and true north corrections provided by [11] for the corresponding datasets. We used the PSF of PDS 70 as a template for the artificial companions and minimized the curvature in the residual image around the planets' positions within a circular aperture with a radius of 80 mas.
To explore the correlation between astrometric and photometric extraction of the companions and to assess the statistical uncertainties of this optimisation, we performed an MCMC sampling using the *MCMCSamplingModule* of *PynPoint*. We used 100 walkers and 200 steps. We excluded the first 20 steps of the chains as the burn-in to the sampling. The presented lower and upper uncertainties represent the 16$^{th}$ and 84$^{th}$ percentiles of the samples, which coincides with 1-sigma confidence levels in the case of a Gaussian posterior distribution.

**Mass determination of PDS 70 c.** The mass of the second companion was estimated by comparing the K-L color and the absolute L magnitude to evolutionary models[41]. To include possible effects of extinction by a circum-planetary disk or the circum-stellar disk, we varied the extinction between 0 and 3 magnitudes. Another strong bias comes from the blending of the circum-stellar disk and the planet in the L-band observations. To estimate this blending we added a conservative 0 to 3 magnitudes to the measured the L-band magnitude to include possible blending effects of the disk. The resulting region in the color-magnitude diagram was overlapping with the 4-12 $M_J$ mass range for the evolutionary model.